\documentclass[twocolumn]{emulateapj}
\slugcomment{Submitted to ApJ}
\shorttitle{Massive Hybrid Stars}
\shortauthors{K. Masuda et al.}
\newcommand{\Slash}[1]{{\ooalign{\hfil$#1$\hfil\crcr\raise.167ex\hbox{/}}}}
\newcommand{\ltsim}{\protect\raisebox{-0.5ex}{$\:\stackrel{\textstyle <}{\sim}\:$}}
\newcommand{\gtsim}{\protect\raisebox{-0.5ex}{$\:\stackrel{\textstyle >}{\sim}\:$}}

\begin{document}
\title{Hadron-Quark Crossover and Massive Hybrid Stars with Strangeness}

\author{Kota Masuda}
\affil{Department of Physics, The University of Tokyo, Tokyo 113-0033, Japan}

\author{Tetsuo Hatsuda}

\affil{Theoretical Research Division, Nishina Center, RIKEN, Wako 351-0198, Japan\\
IPMU, The University of Tokyo, Kashiwa 277-8583, Japan}

\author{Tatsuyuki Takatsuka}
\affil{Iwate University, Morioka 020-8550, Japan}

\begin{abstract}
  Using the idea of 
  smooth crossover from the hadronic matter with hyperons 
  to quark matter with strangeness,  we show that the maximum mass ($M_{\rm max}$) of
  neutron stars with quark matter core can be larger than those without quark matter core.
   This is in contrast to the conventional softening of equation of state
     due to exotic components at high density.  
Essential conditions to reach our conclusion are (i) 
   the crossover takes place at relatively low densities, 
    around 3 times the normal nuclear density,
     and (ii) the quark matter is strongly
   interacting in the crossover region.  By these, 
   the pressure of the system can be greater than that of purely hadronic matter
   at a given baryon density in the crossover density region and  
   leads to $M_{\rm max}$ greater than 2 solar mass.  
   This conclusion is insensitive to the different choice of the 
   hadronic equation of state with hyperons. 
   Several implications of this result to the nuclear incompressibility,
    the hyperon mixing, and the neutrino cooling are also remarked. 
\end{abstract}

\keywords{dense matter, elementary particles, equation of state, stars: neutron}

\section{Introduction}

  Since neutron stars (NSs)
   are gravitationally bound, observation of the NS mass ($M$) provides a precious probe for 
   the equation of state (EOS) and the relevant composition and structure of dense matter.  The larger $M$ requires a stiffer EOS and thereby gives a stringent condition on the state of matter, since various new phases proposed so far, such as pion condensation, kaon condensation, hyperon ($Y$)  mixing, 
  deconfined quark matter and so on, are believed to soften the EOS.  

   For example, the EOS is softened dramatically when $Y$ takes part in NS cores and the theoretical maximum mass  fails to exceed the observed mass, $M_{\rm obs} = 1.44 M_{\odot}$
  for PSR1913+16 with $M_{\odot}$ being the solar mass.
  This contradiction between theory and observation 
   may indicate a missing repulsion in  hyperonic matter such as the 
  3-body interaction (TNI) acting universally  in the nucleon and hyperon sectors 
   (\cite{Nishizaki:2001,Takatsuka:2006,Takatsuka:2008}).

  Very recently, the observation of heavy NS with 
  $M  = (1.97 \pm 0.04)M_\odot$  for PSR J1614-2230 (\cite{Demorest:2010})
   has been reported, which  causes  even stronger  constraint on the EOS.  
   In particular, it stimulates intensive discussions on whether  
 such a heavy NS can have exotic components in the central core (\cite{Ozel:2010,Kurkela:2010,Kim:2011,Klahn:2011,Lattimer:2010,Weissenborn:2011,
  Bonanno:2012,Chen:2012,Schramm:2012,Whittenbury:2012}).
 
  In particular, in many of the previous works treating the 
transition from the hadron phase to the quark phase at high density,
  point-like hadrons and quarks are
   treated as independent degrees of freedom, and Gibbs 
    phase equilibrium conditions are imposed.
   However, 
  quantum chromodynamics (QCD) tells us that such a description
 is not  fully justified, since all hadrons
  are extended objects composed of quarks and gluons. Indeed,
  one may expect a gradual onset of quark degrees of freedom
 in dense matter associated with the percolation of finite-size hadrons,
  i.e., a smooth crossover  from hadronic matter to quark matter:
  See seminal works on hadron percolation  (\cite{Baym:1979,Celik:1980}),
   recent works on hadron-quark crossover (\cite{Baym:2008,Maeda:2009}) 
    and  a recent review on phase transitions in dense QCD (\cite{Fukushima:2011}).    
    
  The aim of this paper is to make a phenomenological study
   on the mechanism realizing  
  hybrid stars (NS with a quark core) compatible with 2$M_{\odot}$ 
  from the point of view of the smooth crossover between 
   the hadronic matter with hyperons and the quark matter with strangeness.
  Unlike the traditional approach to the hadron-quark transition,
   the crossover picture gives us 
   a novel tool to study whether the hybrid  star with a quark core 
  is compatible with massive neutron stars (\cite{Takatsuka:2011}).

\section{Hadronic Matter with Strangeness}

 Let us start with an example of 
 hadronic EOS with hyperons (H-EOS) (\cite{Nishizaki:2001,Nishizaki:2002,Takatsuka:2006})  
 obtained with the following procedure:
(i) Effective two-baryon potentials ${\tilde V}_{BB'}$ ($B= n$, $p$, $\Lambda$, 
$\Sigma^-$) are constructed on the
 basis of the G-matrix formalism to take into account their density-dependence.
(ii) A phenomenological 3-body nucleon force ${\tilde U}_{NN'}$ 
  expressed in a form of two-body potential  (\cite{Friedman:1981}) is introduced
   to reproduce the saturation of symmetric  nuclear matter
    (the saturation density $\rho_0=0.17$ fm$^{-3}$ and the binding 
energy $E_0=-16$ MeV)  and the incompressibility $\kappa$.
(iii) We assume universal three-body repulsion 
 even for the hyperons through the replacement,
  ${\tilde U}_{NN'} \rightarrow {\tilde U}_{BB'}$. 
(iv) By using ${\tilde V}_{BB'}+{\tilde U}_{BB'}$,
 we obtain the hadronic EOS under charge neutrality and $\beta$-equilibrium
  and calculate particle composition $y_i$ ($i =n$, $p$, $\Lambda$, 
$\Sigma^-$, $e^-$ and $\mu^-$) as a function of total baryon density $\rho$.
This gives our H-EOS of neutron star matter with hyperon mixing.

The stiffness of H-EOS is specified by $\kappa$; the case with 
$\kappa=300$ (250) MeV is denoted by TNI3u (TNI2u).  
Here the subscript ``u" means that the 3-body force is introduced universally including 
the hyperon sector.
 The maximum mass ($M_{\rm max}$) of the NSs with hyperon-mixed core 
 with universal 3-body force
  becomes  $M_{\rm max}\simeq1.82(1.52)M_{\odot}$ for 
TNI3u  (TNI2u): On the other hand,   with  only the nucleon 3-body force (TNI3), 
 it reduces to $M_{\rm max}\simeq1.1M_{\odot}$ 
  which is not even compatible with    $M_{\rm obs}=1.44M_{\odot}$.
 One of the effects of the universal 3-body force is to delay the onset of 
  hyperon mixing, e.g.  $\rho_\Lambda \simeq\rho_{\Sigma^-}\simeq4\rho_0$ 
for TNI3u and TNI2u cases, while 
$\rho_{\Sigma^-}\simeq2.2\rho_0$, $\rho_\Lambda \simeq2.5\rho_0$ 
for TNI3.

 To check the model-dependence of the H-EOS, we 
 consider two alternative equations of state with hyperons: 
  AV18+TBF+$\Lambda\Sigma$ (\cite{Baldo:2000}) which is based on 
   the G-matrix approach with the AV18 nucleon-nucleon potential,
  Urbana-type three-body nucleon potential and the Nijmegen
 soft-core nucleon-hyperon potential;
  SCL3$\Lambda\Sigma$ (\cite{Tsubakihara:2010}) which is 
 a relativistic mean-field model  with  chiral SU(3) symmetry.
  The maximum masses (central baryon densities, radius) of NSs composed only by hadronic EOSs, AV18+TBF+$\Lambda\Sigma$ and SCL3$\Lambda\Sigma$, are
 $M_{\rm max}\sim 1.22M_{\odot}$ ($7.35\rho_0$, 10.46km) and $M_{\rm max}\sim 1.36M_{\odot}$ (5.89$\rho_0$, 11.42km),
  respectively.

 In Fig.1, we plot the energy per particle $(E/A)$ of 
 hadronic EOS with hyperons (TNI3u, TNI2u, TNI3, and SCL3$\Lambda\Sigma$)
 together with the   nuclear EOS without hyperons, APR  (\cite{Akmal:1998}).
 We do not show AV18+TBF+$\Lambda\Sigma$  since it is almost the same with TNI3.   
 Filled circles on each line denote the
   density where the hyperons start to mix.  
  From this figure, one can see that (i) the mixture of hyperons softens the equation
   of state relative to APR, and (ii) onset of the hyperon mixture is shifted to 
    higher density if we consider the universal 3-body interaction.     
 In the following, we adopt TNI3u as a typical example of the H-EOS and comment on
  the results with other H-EOSs at the end.
       
\begin{figure}[t]
\epsscale{1.0} 
\plotone{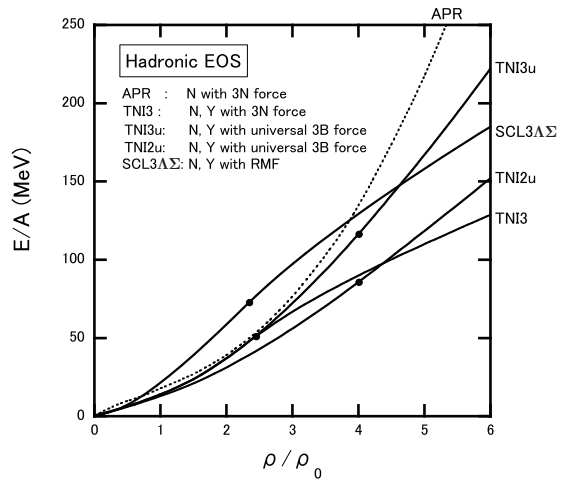} 
\caption {Energy per particle $(E/A)$ as a function of the 
 total baryon density $\rho$. Solid lines denote EOSs with hyperon mixing;
TNI3u (universal 3-baryon force with $\kappa=300$MeV), TNI3 (3-nucleon force with $\kappa=300$MeV), 
TNI2u(universal 3-baryon force with $\kappa=250$MeV),
  SCL3$\Lambda\Sigma$ (relativistic mean-field model  with  chiral SU(3) symmetry).
 Dotted lines denote the EOS without hyperon, APR (AV18+$\delta v$+ UIX$^*$).
  Filled circles on each line show the
   density where the hyperons start to mix.}  
 
\label{fig:fig1}
\end{figure}

\section{Quark Matter with Strangeness}

 Next we consider EOS of quark matter with strangeness (Q-EOS).
 Around the baryon density of a few times $\rho_0$ where the hadron-quark crossover
  is expected to take place at zero temperature, 
  the deconfined (or partially deconfined) quarks would be 
  still strongly interacting.   Analogous 
  situation at high temperature has been expected theoretically and
   evidences of a strongly interacting quark-gluon plasma (sQGP)
   were found experimentally  in relativistic heavy-ion collisions 
    (see e.g. a review, \cite{Fukushima:2011}).

     Since numerical simulations of quantum chromodynamics (QCD)
     on the space-time lattice at high baryon density 
   is not available due to the notorious sign problem, we 
  treat the 
  strongly interacting quark matter (sQM)
   at zero temperature 
 by  the (2+1)-flavor Nambu$-$Jona-Lasinio (NJL) model (see the reviews,
  (\cite{Vogel:1991,Klevansky:1992,Hatsuda:1994,Buballa:2005}). It is an 
 effective theory of QCD and is particularly useful for taking into account the non-perturbative
 phenomena such as the partial restoration of chiral symmetry at high density.  The model
  Lagrangian reads
\begin{eqnarray}
{\cal L}_{\rm NJL}
&=&\overline{q}(i \Slash \partial-m)  q+\frac{G_S}{2}\sum_{a=0}^{8}[(\overline{q}\lambda^a q)^2+(\overline{q}i\gamma_5\lambda^a q)^2]\nonumber \\
& &+ G_D[\mathrm{det}\overline{q}(1+\gamma_5) q+ {\rm h.c.}]
-\frac{g_{_V}}{2}(\overline{q}\gamma^{\mu} q)^2 ,
\label{eq-1}
\end{eqnarray}
where the quark field $q_i$ ($i=u,d,s$) has three colors and  three flavors with
the current quark mass $m_i$.
 The term proportional to $G_S$ is a $U(3)_L \times U(3)_R$ symmetric
 four-fermi interaction where  $\lambda^a$ are the Gell-Mann matrices with
  $\lambda^0=\sqrt{2/3}I$. 
 The  term proportional to $G_D$ is the 
  Kobayashi$-$Maskawa$-$'t Hooft (KMT)
 six-fermi interaction which breaks $U(1)_A$ symmetry. 
 The third term proportional to $g_{_V}$ is a phenomenological vector-type interaction.
  It has some varieties depending on its flavor-structure: Here we use 
 the form given in Eq.(\ref{eq-1}) which 
 leads to an universal flavor-independent repulsion among quarks. 
 
 In the mean-field approximation, the constituent quark masses $M_i$ ($i=u,d,s$) 
 are generated  dynamically through the NJL interactions ($G_{S,D}$), 
$  M_i=m_i-2G_S \sigma_i -2G_D \sigma_j \sigma_k$,
 where $\sigma_i = \langle \bar{q}_iq_i \rangle$
 is the quark condensate in each flavor, and ($i$, $j$, $k$) corresponds to  the cyclic 
 permutation of $u, d$ and $s$.  On the other hand,
 the vector interaction ($g_{_V}$) leads to an effective
  chemical potential (\cite{Asakawa:1989}), 
$\mu^{\rm (eff)}_i = \mu_i - g_{_V} \sum_i \langle q_i^{\dagger} q_i \rangle$.

 Basic parameters of the NJL model are determined from hadron phenomenology:  In this 
 paper, we adopt so-called the HK parameter set (\cite{Hatsuda:1994}),    
$\Lambda=631.4$ MeV, $ G_S\Lambda^2=1.835$, $G_D\Lambda^5=9.29$, 
$m_{u,d}=5.5$ MeV, $m_s=135.7$ MeV, 
where $\Lambda$ is the three-momentum cutoff.
 The magnitude of $g_{_V}$ has not been well determined,
  but recent studies  of the model applied to the QCD phase diagram suggest that it can be   
 comparable to or even larger than  $G_S$ (\cite{Bratovic:2012,Lourenco:2012}),
  so that we change its value in the range, 
\begin{eqnarray} 
  0 \le \frac{g_{_V}}{G_S} \le 1.5 .
  \label{eq:GV}
\end{eqnarray}
 The EOS of strongly interacting quark matter with strangeness is
 obtained from the above model under charge neutrality and $\beta$-equilibrium
  with $u$, $d$, $s$, $e^-$ and $\mu^-$.  It turns out that the 
  strange quarks  starts to appear  at  $\rho \simeq 4\rho_0$ regardless of the 
   magnitude of $g_{_V}$.  Also,  $\mu^-$ never appears under the presence of the strangeness.

\section{Hadron-Quark Crossover}

To realize the smooth transition between the hadronic matter
 and the  quark matter, we first define a ``crossover density
 region" characterized by
  its central value $\bar{\rho}$
  and its width  $\Gamma$.
   The description of the matter in terms of the pure hadronic EOS
  is accurate for $\rho \ll \bar{\rho} - \Gamma$, while the 
  description in terms of pure quark EOS is accurate for  
  $\rho \gg \bar{\rho} +  \Gamma$.
 In the crossover region, $ \bar{\rho} - \Gamma \ltsim \rho \ltsim  \bar{\rho} + \Gamma$,
  both hadrons and quarks are strongly interacting, so that
 neither pure hadronic EOS nor pure quark EOS are reliable.
 Under this situation,  we make a  
  smooth interpolation of the  two descriptions  
 phenomenologically 
 with a procedure similar to the one at high temperature   (\cite{Asakawa:1997}):
\begin{eqnarray}
P &=&P_H\times f_-+P_Q\times f_+, 
\label{eq:HQ-EOS-0} \\
f_{\pm} &=& \frac{1}{2} \left( 1\pm \mathrm{tanh}\left( \frac{\rho-\bar{\rho}}{\Gamma} \right) \right),
\label{eq:HQ-EOS}
\end{eqnarray}
where   $P_{H}$ and $P_Q$ are  the pressure in the pure hadron matter and that in  
the pure quark matter, respectively.  
 One should not confuse our interpolated pressure in Eq.(\ref{eq:HQ-EOS-0}) with that 
 of  the hadron-quark mixed phase associated with the first order phase transition:
  In our crossover picture,  we are free from the 
 `phase equilibrium condition' such as  $P_Q=P_H$ at fixed chemical potential.
  (For phenomenological attempts to interpolate H-EOS and Q-EOS within the
  conventional picture of 
  first order phase transition, see e.g. \cite{Burgio:2002} and references therein.)

 The energy density $\varepsilon$ as a function of $\rho$ is obtained by
  integrating the thermodynamical relation,  
$  P=\rho^2 {\partial(\varepsilon/\rho)}/{\partial \rho}$,
   which guarantees the thermodynamical consistency for any values of 
   $\bar{\rho}$ and $\Gamma$.            
  To explore the relation between the interpolated EOS and the maximum mass of 
   neutron stars, we change the parameters  ($\bar{\rho}$, $\Gamma$) under
  two conditions: (i) The system is always 
 thermodynamically stable $dP/d\rho >0$, and 
 (ii) the normal nuclear matter is described dominantly by $P_H$, i.e.
  the condition   $ \bar{\rho} -2 \Gamma > \rho_0 $.

\begin{figure}[!t]
\epsscale{0.9} 
\plotone{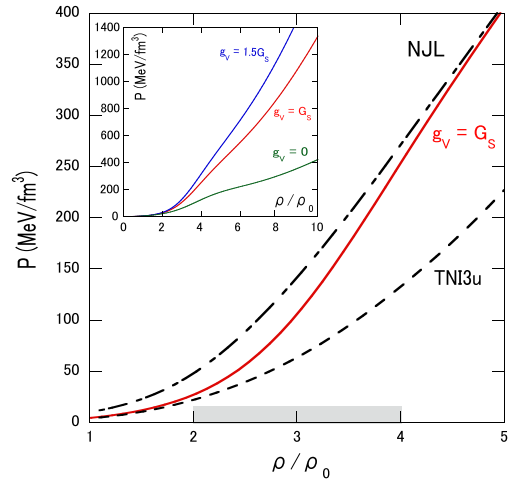} 
\caption{Solid line: 
Pressure ($P$) as a function of baryon density 
$\rho$ obtained by interpolating H-EOS and Q-EOS with $g_v=G_s$
 in the crossover  region (the shaded region in density).
  Dash-dotted line: Q-EOS  in the NJL model with $g_{_V}=G_S$.
  Dashed line: H-EOS  with the TNI3u interaction.
     The inset shows EOSs interpolated with 
  $g_{_V}/G_{S}=0, 1.0, 1.5$. }
\label{fig:fig2}
\end{figure}

\begin{figure}[!h]
\epsscale{0.9} 
\plotone{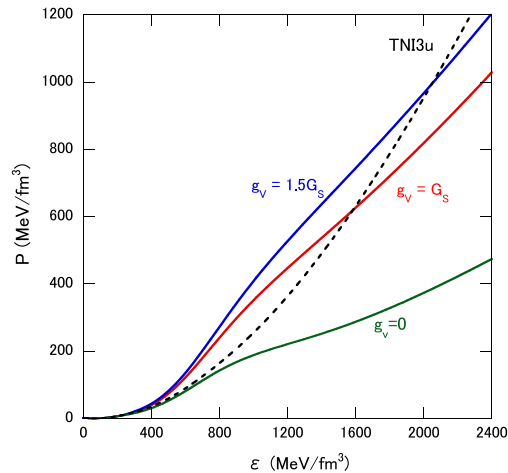} 
\caption{Solid lines: $P$ vs. energy-density ($\varepsilon$) for  
 the interpolated EOSs  with $g_{_V}/G_S=0,1.0,1.5$. Dashed line denotes hadronic matter
  with TNI3u interaction.}
\label{fig:fig3}
\end{figure}

 Shown in Fig.~\ref{fig:fig2} is the pressure $P$ as a function of baryon density $\rho$
  obtained by interpolating  TNI3u H-EOS (the dashed line)  and NJL Q-EOS (the dash-dotted line)
 with $g_{_V} = G_S$ according to the procedure in Eq.(\ref{eq:HQ-EOS}).
 (Adopting TNI2u instead of TNI3u gives essentially the same final result.)
  As indicated by the  shaded region on the horizontal axis,  
  crossover parameters  in this figure are 
   $(\bar{\rho}, \Gamma) = (3\rho_0, \rho_0)$, which leads
  to a smooth interpolation between the relatively soft H-EOS and the relatively stiff Q-EOS.
  Note that we have  $ P_{H} < P < P_Q$  in the crossover region, 
  which will never be realized in the standard Maxwell or Gibbs construction 
  assuming  the first-order transition, where  $P < P_H$ always
   holds in the mixed phase.
  As will see later, the stiff EOS at $(2-4)\rho_0$ is most important in  
  obtaining NSs with mass greater than 2$M_{\odot}$.

  In the inset of  Fig.~\ref{fig:fig2}, we show
  EOSs interpolated in the same crossover region
  but with  different values of the
  vector coupling ($g_{_V}/G_S=0, 1.0, 1.5$ motivated by Eq.(\ref{eq:GV})).
  As $g_{_V}$ increases,
  EOS becomes stiffer at high density, while the density where
  the strange-quark starts to appear ($\rho \simeq 4\rho_0$) is
  insensitive to $g_{_V}$.

  In Fig.~{\ref{fig:fig3}, we plot the pressure $P$ as a function of the 
  energy density  $\varepsilon$ for the three EOSs  interpolated
  in the crossover  region, $(\bar{\rho}, \Gamma) = (3\rho_0, \rho_0)$.
  These curves are the basic inputs when we solve the Tolman-Oppenheimer-Volkov (TOV) equation.
     The case for H-EOS with TNI3u is also plotted by the dashed line for comparison.
   Existence of the region where $P$ (the solid line) becomes
    larger than $P_{H}$ (the dashed line) is essential for having large 
     $M_{\rm max}$ as shown below.

\section{Hybrid star structure}

In Fig.~\ref{fig:fig4}, the solid lines indicate
 NS masses as a function of the central density $\rho_c$ 
 obtained by solving the TOV equation with the EOSs given in the inset of 
 Fig.~\ref{fig:fig2}, i.e., interpolated EOSs between  TNI3u H-EOS and NJL Q-EOS
  with $g_{_V}/G_S=0,1.0,1.5$.  
 Results with the hadronic EOS only  (TNI3u, AV18+TBF+$\Lambda \Sigma$ and SCL3$\Lambda \Sigma$) are also plotted for comparison by the dashed lines. 
The filled circle indicates a point beyond which the 
  strangeness appears in the central core of the star. 
 The cross symbols indicate the points where $M_{\rm max}$
 is achieved.

\begin{figure}[t]
\epsscale{0.9} 
\plotone{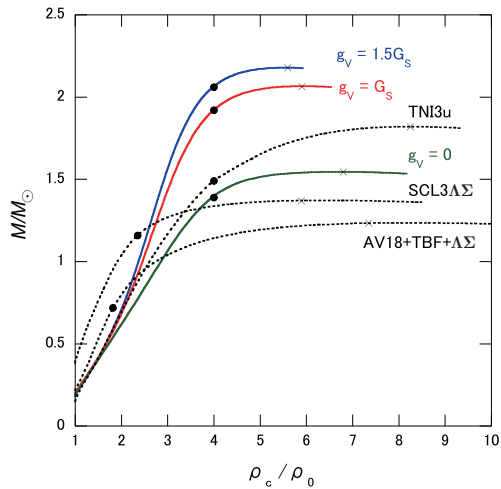} 
\caption{ Solid lines: NS mass ($M$) vs. the central baryon density ($\rho_c$) obtained from
 interpolated EOSs between TNI3u H-EOS and NJL Q-EOS with $g_{_V}/G_S=0, 1.0, 1.5$. 
Dashed lines: The same quantities for H-EOS with TNI3u, AV18+TBF+$\Lambda \Sigma$ and SCL3$\Lambda \Sigma$.
   The cross symbols denote the points of $M_{\rm max}$, while 
   the filled circles denote the points beyond which the strangeness appears. }
\label{fig:fig4}
\end{figure}

\begin{figure}[t]
\epsscale{0.9} 
\plotone{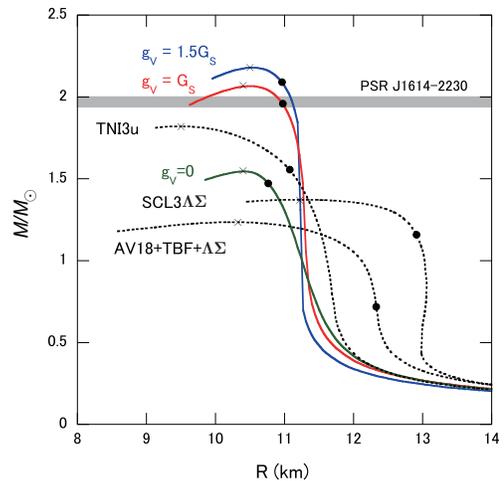} 
\caption{ Solid lines: $M$-$R$ relation of the NSs with the interpolated EOSs between TNI3u H-EOS
and NJL Q-EOS for $g_{_V}/G_S=0, 1.0, 1.5$.
 Dashed lines: The same quantities for H-EOSs with TNI3u, AV18+TBF+$\Lambda \Sigma$ and SCL3$\Lambda \Sigma$.
  The cross symbols denote the points where $M$ reaches $M_{\rm max}$.
   The filled circles denote the point beyond which the strangeness appears.
    The gray band denotes $M=(1.97\pm0.04)M_{\odot}$ for PSR J1614-2230 (\cite{Demorest:2010}.)}
\label{fig:fig5}
\end{figure}

  For TNI3u H-EOS only, we have $M_{\rm max}=1.82 M_{\odot}$, while, for interpolated EOS 
 with  $g_{_V}/G_S \gtsim 1$, we have  $M_{\rm max} >2 M_{\odot}$.  
  This indicates that the existence of the quark core inside neutron stars
  is compatible with the observation of high-mass neutron star as long as (i) there is a 
  smooth crossover between the hadronic phase and the quark phase, and (ii) there
  exits a strong quark correlations with repulsive nature inside the quark matter.

In Fig.~\ref{fig:fig5}, the solid lines indicate the mass($M$)-radius($R$) relation of the NSs 
 with the interpolated EOSs given in the inset of Fig.~\ref{fig:fig2}.
 Results with the H-EOS only  (TNI3u, AV18+TBF+$\Lambda \Sigma$ and SCL3$\Lambda \Sigma$)
  are also plotted for comparison by the dashed lines.
 The gray band shows  $M  = (1.97 \pm 0.04)M_\odot$  corresponding to
  PSR J1614-2230 (\cite{Demorest:2010}).
    Because of the stiff EOS at high density  for $g_{_V}/G_S \gtsim 1$, 
  the radius $R$ for  $0.5 \ltsim M/M_{\odot} \ltsim 2$ takes almost a constant value,
  11$-$11.5 km(\cite{Ozel:2011}).  
  Also,  strangeness emerges only for massive stars with $M \gtsim 2 M_{\odot}$.
 
\begin{table}[!h]
\caption{$M_{\rm max}/M_{\odot}$ and $\rho_c/\rho_0$ in the parenthesis
  under the variation of the crossover region, $\bar{\rho}$ and $\Gamma$.} 
\begin{center}
  \begin{tabular}[c]{c|c|c|c|c}  \hline \hline
 & \multicolumn{2}{|c|}{${\Gamma}/{\rho_0}=1$} & \multicolumn{2}{|c}{${\Gamma}/{\rho_0}=2$} \\ 
 \cline{2-3}  \cline{3-4}\cline{4-5}
   $\bar{\rho}$  
 & ${g_{_V}}$=${G_S}$ 
 & ${g_{_V}}$=$1.5{G_S}$ 
 & ${g_{_V}}$=${G_S}$ 
 & ${g_{_V}}$=$1.5{G_S}$   \\ \hline 
 $3\rho_0$   & 2.07 (5.9) & 2.18 (5.6) & $-$ & $-$    \\ 
 $4\rho_0$   & 1.93 (6.7) & 2.00 (6.6) & $-$ & $-$    \\ 
 $5\rho_0$   & 1.79 (7.7) & 1.83 (7.4) &1.82 (7.4)&1.86 (7.3)\\ 
 $6\rho_0$   & 1.70 (8.3) & 1.70 (8.3) &1.73 (8.0) &1.74 (8.0)\\
\hline \hline
\end{tabular} 
\end{center}
\end{table}

So far, we have taken $(\bar{\rho}, \Gamma)=(3\rho_0, \rho_0)$ as 
characteristic parameters  for the crossover  region. 
In Table I, we show how $M_{\rm max}$ and $\rho_c$ depend
on the choice of these parameters for fixed values of $g_{_V}$.
 As the crossover  region becomes lower
  and/or wider in baryon  density,  the EOS  
  characteristic to the NS core ($\rho = (2-4) \rho_0$)  becomes stiffer
   This is why
  $M_{\rm max} > 2M_{\odot}$ becomes possible for $\bar{\rho}/\rho_0 =3$ and $4$.

 \begin{table}[!h]
\caption{$M_{\rm max}/M_{\odot}$ and $\rho_c/\rho_0$ in the parenthesis
  for different H-EOSs with  $\bar{\rho}=3\rho_0$ and $\Gamma/\rho_0=1$.} 
\begin{center}
  \begin{tabular}[c]{c|c|c|c}  \hline \hline
 & \multicolumn{3}{|c}{${\Gamma}/{\rho_0}=1$}  \\ 
 \cline{2-3}\cline{3-4} 
   H-EOS
 & ${g_{_V}}$=$0$ 
 & ${g_{_V}}$=${G_S}$
 & ${g_{_V}}$=$1.5{G_S}$ \\ \hline 
 TNI3u &1.55 (6.8) &2.07 (5.9) & 2.18 (5.6)     \\ 
 TNI2u &1.50 (7.4) &2.05 (6.1) & 2.17 (5.5)     \\ 
 TNI3 & 1.51 (6.8) &2.04 (6.1) & 2.16 (5.5)     \\ 
 AV$_{18}$+TBF+$\Lambda \Sigma$ & 1.52 (6.8) & 2.06 (6.1) & 2.17 (5.5)     \\  
 SCL3$\Lambda \Sigma$& 1.55 (6.8)& 2.06 (5.9) & 2.17 (5.5)     \\ 
\hline \hline
\end{tabular} 
\end{center}
\end{table}

 Now, we examine how $M_{\rm max}$  depends on the different choice of H-EOS.
 Shown in Table 2 is the maximum and central density
   for different H-EOSs with  $\bar{\rho}=3\rho_0$ and $\Gamma/\rho_0=1$.
 As long as $M_{\rm max}$  is concerned, the result is insensitive to
  the choice of the hadronic equation of state with hyperons.

\section{Summary and Concluding Remarks}

 In this paper, we have investigated the maximum mass of hybrid stars with the EOS constructed on the 
 basis of the idea of hadron-quark crossover; hadronic EOS (with hyperons and universal three-baryon repulsion)
  and the quark EOS (with  strangeness and universal two-quark repulsion)
   are interpolated to model the crossover region. 
 
We have found that massive hybrid stars compatible with  $M > 2M_{\odot}$
  are possible as long as the crossover proceeds through relatively low density region, 
 e.g., $\bar{\rho}=(3-4)\rho_0$ with $\Gamma=\rho_0$, and, at the same time, the 
  quark matter is strongly interacting $g_{_V}/G_S \sim 1 - 1.5$.
  By increasing $g_{_V}$ further,  one can even obtain $M_{\rm max}$ close to 2.4 $M_{\odot}$.
   This conclusion is in remarkable contrast with the conventional EOS for 
   hybrid star derived through the Maxwell or Gibbs construction where  
 the resultant EOS becomes always softer than hadronic EOS  and thereby leads to smaller
  $M_{\rm max}$.

  Although we have mainly focused
   on the results using TNI3u H-EOS corresponding to $\kappa=300$MeV,
   very similar results are obtained for TNI2u as shown in Table 2.
  This means an interesting possibility that we can reconcile the existence of massive neutron  stars ($M > 2 M_{\odot}$) with
  the experimental nuclear incompressibility $\kappa=(240\pm20)$MeV  (\cite{Shlomo:2006}).
  We note that  TNI3,  AV18+TBF+$\Lambda\Sigma$ and SCL3$\Lambda\Sigma$ give
  almost the same $M_{\rm max}$ with that of TNI3u and TNI2u.
  This does not, however, imply the irrelevance of the universal three-body force:
   For example, the interpolated EOS with H-EOS (TNI3u) and  Q-EOS ($g_v=G_s$) 
   leads to  strangeness mixing only above 4$\rho_0$ due to the universal three-body force, 
  so that the hybrid stars lighter than 1.92$M_{\odot}$  
  do not have strangeness (see Fig.4). Then, the hyperon direct Urca process 
   (e.g., $\Lambda\rightarrow p+e^-+\bar{\nu}_e$, $p+e^- \rightarrow \Lambda +\nu_e$),
  which leads to extremely rapid neutrino-cooling and  contradicts  observations
   (\cite{Takatsuka:2006}), does not take place except for massive NSs.

 Our massive hybrid stars with $M > 2M_{\odot}$ 
    originate from the transition from the soft EOS to stiff EOS
    in the density region  $\rho \sim (2-4)\rho_0$ due to the hadron-quark crossover. 
   Therefore it would be important to explore such crossover by independent
    laboratory experiments with medium-energy heavy-ion collisions.

  Finally, we remark that the crossover region may have 
   rich non-perturbative phases such as 
    color superconducting phases,  inhomogeneous phases, the quarkyonic phase and so on
    (\cite{Fukushima:2011}).  How these structures affect the 
     basic conclusion of the present  paper would be an interesting future problem to be 
     examined. 
 
\acknowledgements
K.M. and T.H. thank Wolfram Weise for discussions.
T.T. thanks Ryozo Tamagaki, Toshitaka Tatsumi and Shigeru Nishizaki for discussions and interests in this work. We also thank K. Tsubakihara and A. Ohnishi for providing us with the numerical
 data of the SCL3 EOS.
This  research was  supported in  part by  MEXT Grant-in-Aid  for Scientific
Research  on  Innovative  Areas(No.2004:20105003) by JSPS
Grant-in-Aid  for Scientific   Research (B)  No.22340052,
and by RIKEN 2012 Strategic Programs for R \& D.\\


\end{document}